\documentclass[%
 reprint,
 amsmath,amssymb,
 aps,
 prd
floatfix,
showkeys
]{revtex4-1}

\usepackage[USenglish]{babel}
\usepackage{slashed}
\usepackage{graphicx}
\usepackage{dcolumn}
\usepackage{bm}
\usepackage{amssymb}
\usepackage{amsmath}
\usepackage{graphicx}
\usepackage[bottom]{footmisc}
\usepackage{epsfig}
\usepackage[utf8]{inputenc} 
\usepackage{ucs} 
\usepackage{amsmath} 
\usepackage{array}
\usepackage{float}
\usepackage{color}
\usepackage[usenames,dvipsnames]{xcolor}
\usepackage{epstopdf}
\usepackage{natbib}
\usepackage{graphicx}
\usepackage{lineno}
\usepackage{slashed}
\usepackage{color}
\usepackage{hyperref}
\usepackage{float}
\usepackage{mathtools,slashed}

\newcommand{\be}{\begin{equation}}
\newcommand{\ee}{\end{equation}}
\newcommand{\bea}{\begin{eqnarray}}
\newcommand{\eea}{\end{eqnarray}}
\newcommand{\beas}{\begin{eqnarray*}}
\newcommand{\eeas}{\end{eqnarray*}}

\newcommand{\bse}{\begin{subequations}}
\newcommand{\ese}{\end{subequations}}

\begin{document}

\title{\bf Time resolution studies for scintillating plastics coupled to silicon photo-multipliers}

\author{Mauricio Alvarado$^1$, Alejandro Ayala$^{1,2}$, Marco Alberto Ayala-Torres$^3$, Wolfgang Bietenholz$^{1}$, Isabel Dominguez$^4$, Marcos Fontaine$^3$, P. Gonz\'alez-Zamora$^5$, Luis Manuel Monta\~no$^3$, E. Moreno Barbosa$^6$, Miguel Enrique Pati\~no Salazar$^1$, V. Z. Reyna Ortiz$^6$, M. Rodr\'iguez Cahuantzi$^6$, G. Tejeda Mu\~noz$^6$, Maria Elena Tejeda-Yeomans$^7$, Luis Valenzuela-C\'azares$^8$, C. H. Zepeda Fern\'andez$^{1,6}$}
\email{Corresponding author, email:hzepeda@fis.cinvestav.mx}
\address{
$^1$Instituto de Ciencias Nucleares, Universidad Nacional Aut\'onoma de M\'exico, Apartado Postal 70-543, CdMx 04510, Mexico\\
$^2$Centre for Theoretical and Mathematical Physics, and Department of Physics, University of Cape Town, Rondebosch 7700, South Africa\\
$^3$Centro de Investigaci\'on y Estudios Avanzados del IPN, Apartado Postal 14-740, CdMx 07000, Mexico\\
$^4$Facultad de Ciencias F\'isico-Matem\'aticas, Universidad Aut\'onoma de Sinaloa, Av. de las
Am\'ericas y Blvd.\ Universitarios, Cd.\ Universitaria, C.P. 80000, Culiac\'an, Sinaloa, Mexico\\
$^5$Departamento de Ingeniería Industrial, Universidad de Sonora, Avenida Universidad S/N, Edificio 5K, Planta Baja, Colonia Centro, 83000  Hermosillo, Sonora, Mexico\\
$^6$Facultad de Ciencias F\'isico Matem\'aticas, Benem\'erita Universidad Aut\'onoma de Puebla, Av. San Claudio y 18 Sur, Edif. EMA3-231, Ciudad Universitaria 72570, Puebla, Mexico\\
$^7$Facultad de Ciencias-CUICBAS, Universidad de Colima, Bernal D\'iaz del Castillo No.\ 340, Colonia Villas San Sebasti\'an, 28045 Colima, Mexico\\
$^8$Departamento de Investigaci\'on en F\1sica, Universidad de Sonora, Boulevard Luis Encinas J.\ y Rosales, Colonia Centro, Hermosillo, Sonora 83000, Mexico}

\begin{abstract}

We present results for time resolution studies performed on three different scintillating plastics and two silicon photo-multipliers. These studies are intended to determine whether scintillating plastic/silicon photo-multiplier systems can be employed to provide a fast trigger signal for NICA's Multi Purpose Detector (MPD). Our results show that such a system made of cells with transverse dimensions of order of a few cm, coupled to silicon photo-multipliers, provides a time resolution of about 50~ps, which can be even further improved to attain the MPD trigger requirements of 20~ps.\\

\end{abstract}


\maketitle

\section{Introduction}\label{I}
Scintillating detectors are used for a wide range of purposes, from medical imaging and radiation security to nuclear and high-energy physics.
Scintillating detectors are commonly coupled to photo-multiplier tubes (PMTs) that can serve as the readout part of the detectors. However, in recent times there has been great interest in the use of silicon photo-multipliers (SiPMs) as an alternative to PMTs, due to their high photon detection efficiency and good time resolution. 
SiPMs are low cost devices which have the advantage of being compact and magnetic field resistant, see e.g. Ref.~\cite{SiPMs}.

Detectors made of scintillating material with SiPMs readout are also promising to obtain a competitive time resolution~\cite{paolo}. These systems are used in high-energy experiments mainly to detect charged particles. The common choice for the scintillation material is either a crystal or a plastic. These materials produce light when subatomic charged particles traverse them. In general, the emitted light has a frequency in the high end of the visible region. The produced photons are collected by the SiPMs that in turn generate electrical signals by means of photoelectric processes. These systems have a distinctive time resolution that makes them
ideal to be used as part of wake-up triggers in high energy collider
experiments. 

One such experiment, which is currently under construction, is the Multi Purpose Detector (MPD) of the Nuclotron-based Ion Collider fAcility
(NICA) at the Joint Institute for Nuclear Research (JINR) \cite{jinr}. The main goal of the MPD is to study the properties of strongly interacting matter at high  temperature and high baryon density~\cite{NICA} and in particular to locate the possible critical end point in the QCD phase diagram~\cite{Ayala}. For these purposes, NICA will provide a wide range of hadron beams, including heavy and light nuclei as well as protons with center of mass energies between 4 and 11 GeV for Au+Au collisions and up to 27 GeV for p+p collisions. 

The design of the MPD includes, as its detector systems, the Inner Tracker (IT), the Time Projection Chamber (TPC), the Zero Degree Calorimeter (ZDC), the Electromagnetic Calorimeter (ECal), the Time of Flight (TOF) and the Fast Forward Detector (FFD)~\cite{mpd,strangenica}. A magnetic field of 0.5 T will be provided by a Magnetic Solenoid. 
The barrel region is designed to cover a pseudorapidity range $|\eta|\leq$ 1.2 with the end caps and the FFD extending the coverage to 1.2 $\leq |\eta|\leq$ 2 and 3 $\leq|\eta|\leq$ 3.9, respectively. The TPC, ECal, TOF, ZDC, FFD and the Magnet are the detector systems planned to be included in the first run, which is scheduled for 2020. The IT will be added in a second run. Two Forward Spectrometers (FS-A and FS-B) based on toroidal magnets, may be installed for the third run and they are planned to cover the region   2 $\leq|\eta|\leq$ 4. Given the absence of the IT for the first run, the FFD is supposed to provide the trigger for the  TOF. Simulations show that for the large multiplicity environment in Au+Au collisions, the FFD provides a good time resolution to serve as such a trigger. This may not be the case, however, for the lower multiplicities produced in collisions of lighter systems~\cite{FFD}. 

The need to find an adequate trigger for the TOF has prompted research with the purpose to design a fast detector component which may also serve as a redundant element and perhaps to extend the rapidity coverage of the originally planned MPD elements. A possibility for such a detector element has been put forward in Ref.~\cite{NIM}. This is a beam-beam counter consisting of two discs made of hexagonal scintillating plastic tiles. Each disc is planned to be located 2 m away on either side of the interaction point along the beam pipe. Details of its design and capabilities can be found in Ref.~\cite{NIM}, which also shows the time resolution for one kind of plastic tile coupled to either a PMT or a SiPM, obtained in a pion beam test made at the T10 facility at CERN.
\begin{figure}[t!]
\begin{center}
\includegraphics[width=0.3\textwidth]{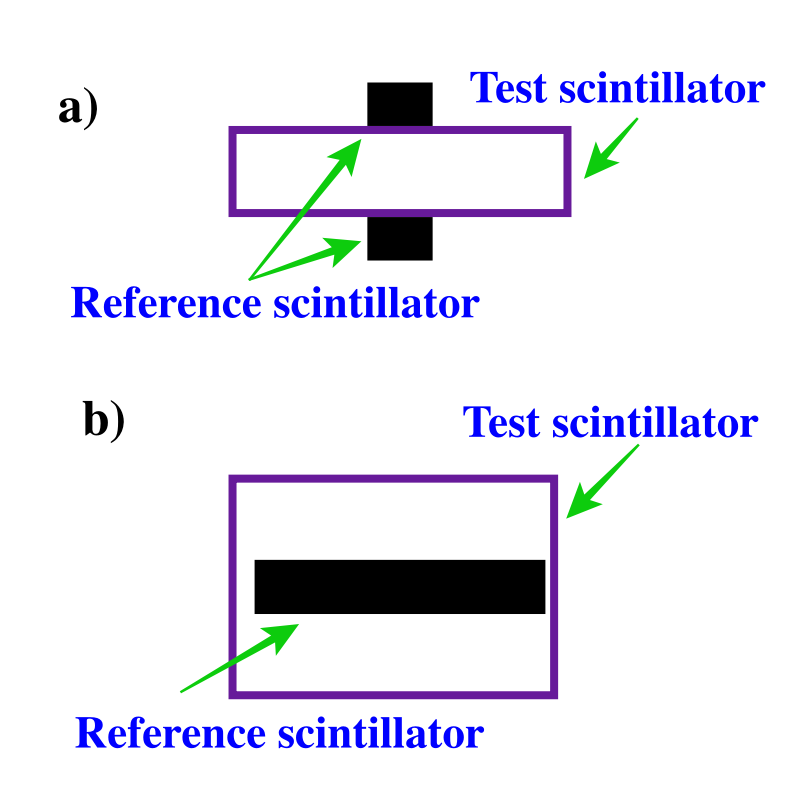}
\end{center}
\caption{Telescope arrangement. The figure shows the representation of the test scintillating plastic and reference scintillating plastic (BC408). (a) side view. (b) front view.}
\label{arrange}
\end{figure}

In order to complement these time resolution studies, here we present the results of laboratory tests using a telescope array triggered both by cosmic rays and charged particles coming from radiation sources, to measure the time resolution of several kinds of plastic tiles coupled to different SiPMs.  We show that a time resolution similar to the one obtained in the pion beam test is achieved. 

This work is organized as follows: In Sec.~\ref{2} we describe the experimental set up. In Sec.~\ref{3} we outline the way we carried out the data acquisition and present the analysis of the results. In Sec.~\ref{4} we discuss these results and present our conclusions. 

\section{Experimental set up}\label{2}

The study was performed using tiles made of two different  scintillating commercial plastics: BC404\textregistered Bicron, BC408\textregistered Bicron. In addition, we also tested one scintillating plastic developed at {\it Facultad de Ciencias F\'{\i}sico-Matem\'aticas, Benem\'erita Universidad Aut\'onoma de Puebla (BUAP), M\'exico}~\cite{scinpuebla} (BUAP-SP). We tested the performance of two different SiPMs:   (C-60035-4P-EVB)\textregistered SensL \cite{sensl} and
(S10985-050C)\textregistered Hamamatsu \cite{hamma}, which were coupled to the scintillating plastics to compare their light acquisition and time response performance.  We placed these SiPMs at different locations over each plastic surface in order to identify the optimal geometrical configuration for the signal acquisition. We also implemented some improvements in the data readout by testing different configurations for the electronics that send the light sensor signals to the acquisition set up. The improvements were meant to obtain a low noise fast signal (reaching a peak time in less than 30~ns) using common electronic devices and materials. The BC408 and BC404 where taken as square tiles with dimensions $10\times10\times2$~cm$^3$, whereas for the BUAP-SP we use a hexagon tile 5.6 cm high and 1.5 cm thick. To test the performance of the three scintillating plastic tiles, we used a telescope arrangement consisting of two BC408 scintillating plastics as the time references ($5\times2\times1$~cm$^3$). This arrangement is depicted in Fig.~\ref{arrange}. The telescope arrangement is used as the trigger that selects events hitting the two references and the tested scintillating plastics within a time window of 20~ns. The charged particles passing through the array were either cosmic rays or they came from radioactive sources. For the latter we used Na-22, Mn-54, Co-57, Fe-55, Co-60, Cs-137, Ba-133 and Cd-109, in order to choose the optimal source that provided the best statistics.

\section{Analysis}\label{3}
\begin{figure}[htb]
\centering
\includegraphics[width=0.48\textwidth]{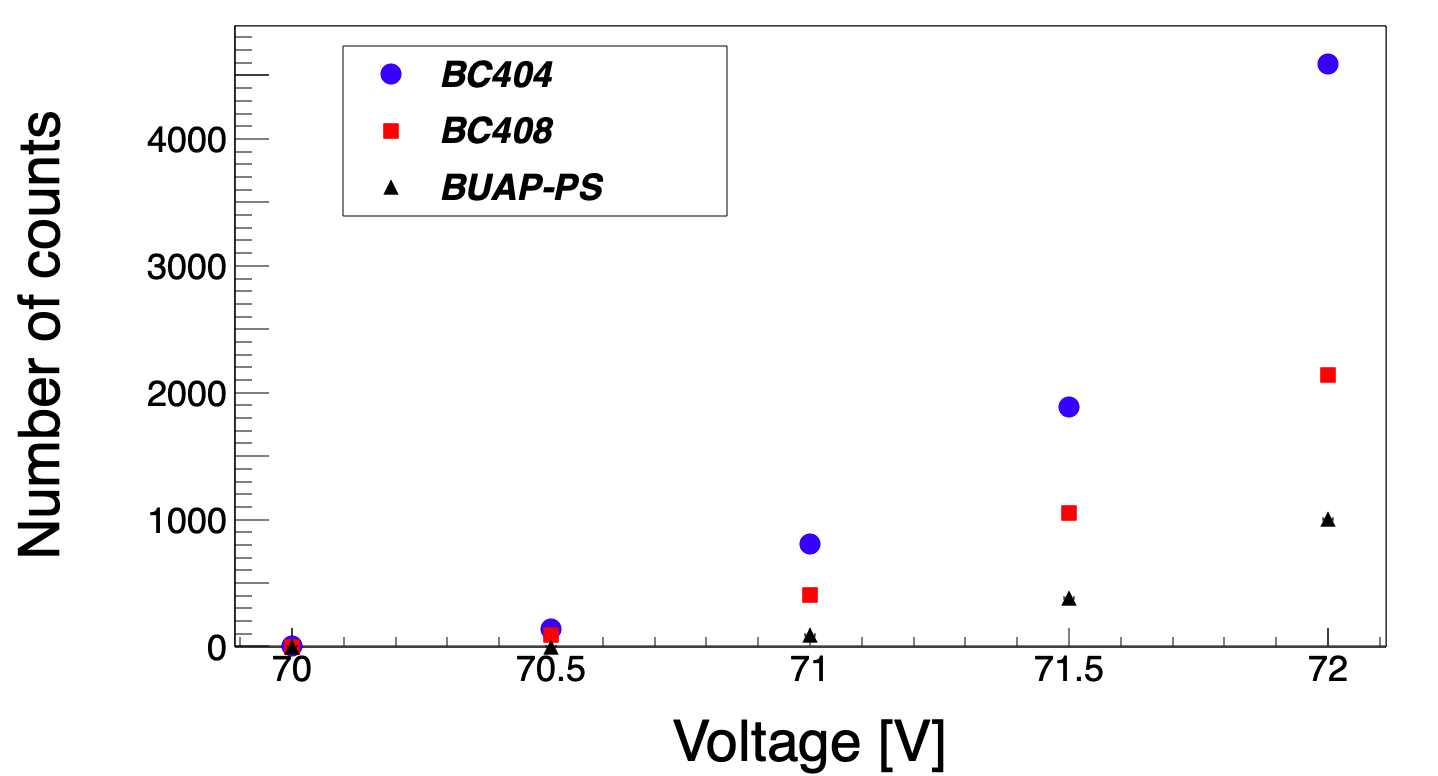}
\caption{Comparison between the three tested scintillating plastics. BC404 achieved the best light acquisition performance for the adequate bias voltage of the Hamamatsu light sensor.}\label{compa}
\end{figure}

In order to test the performance of the three scintillating plastics, we first connected the Hamamatsu SiPM using a bias around 72~V. We tested the different radioactive sources and chose to use Na-22 that provides the highest number of counts. The radioactive source was located at the center of each scintillating plastic. We increased the voltage from 70~V to 72~V in steps of 0.5~V. This range corresponds to the SiPM operating voltage. For the electronics, we used a logic unit (Lecroy 365AL) and a Scaler (CAEN N1145). For each operating voltage we allowed five minutes for the light acquisition time. We observed that the BC404 is the most efficient material for the operating conditions. The comparison between the performance of our three scintillating plastics is shown in Fig.~\ref{compa}, from where we infer that the BC404 achieved the best light acquisition performance for the optimal bias voltage of the Hamamatsu light sensor.

\begin{figure}[htb]
\centering
\includegraphics[width=0.45\textwidth]{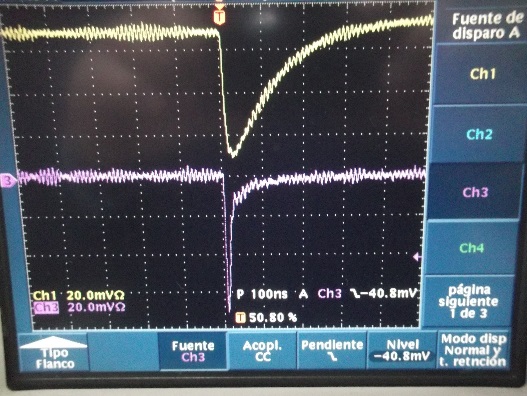}
\caption{Comparison between the output signal from the light sensor before (upper signal) and after (lower signal) the electronic circuit modification.}
\label{signal}
\end{figure}

\begin{figure}[htb]
\centering
\includegraphics[width=0.47\textwidth]{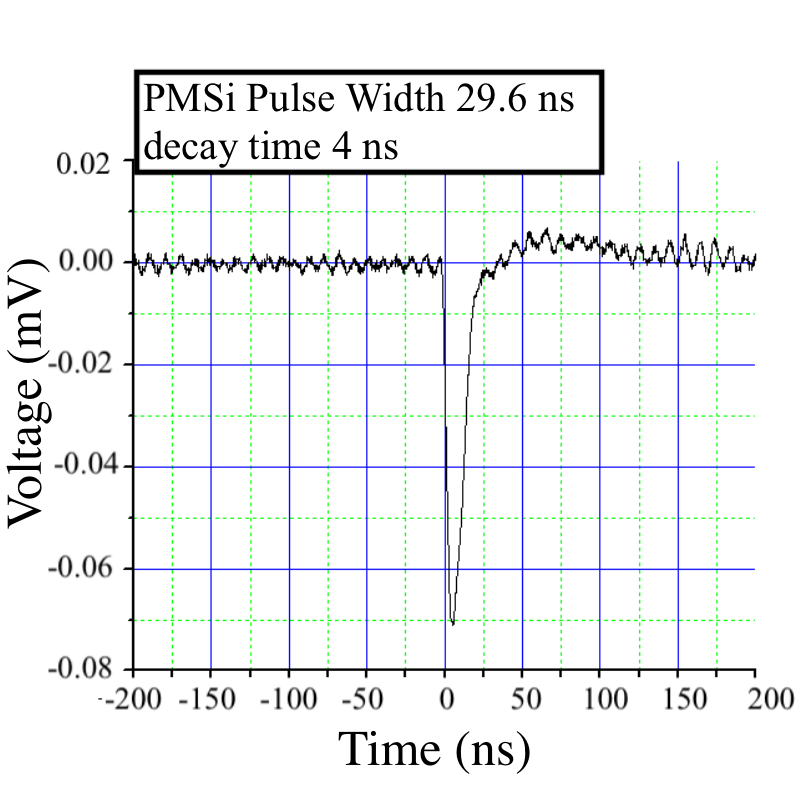}
\includegraphics[width=0.48\textwidth]{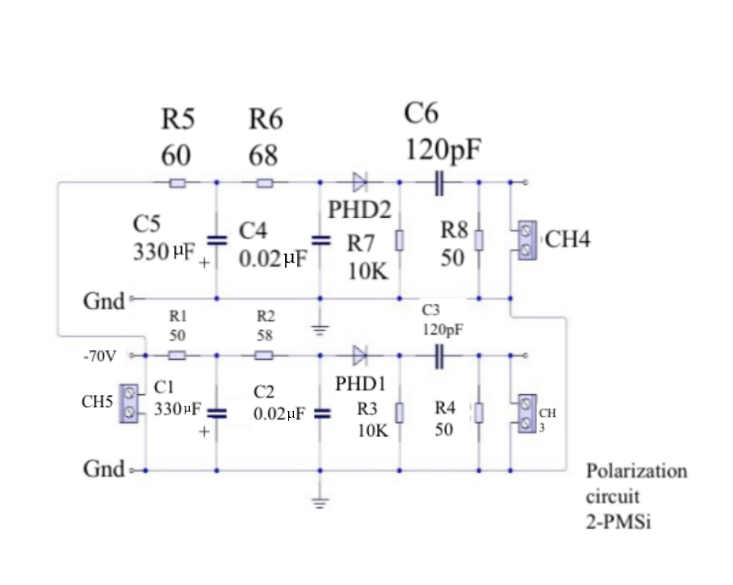}
\caption{(Top) Further improvements to the shape of the signal. (Bottom) Scheme of the circuit used to bias the light sensors, where R, C, CH, Gnd and PHD stand for Resistance, Capacitor, Channel, Ground and Photo-diode, respectively.}\label{electronic}
\end{figure}

The output signal through the bias circuit had the original shape shown by the upper curve in Fig.~\ref{signal}. 
This signal had a recovery time of around 200~ns. After modifying the circuit, we obtained a signal having a peak time around 5~ns, and a recovery time of less than 30~ns. This is shown by the lower curve in Fig.~\ref{signal}. The improvement consists of adding a low value capacitor and a $50~\Omega$ resistance to the bias circuit. The modification had the purpose to attain the same impedance for both, the bias circuit's output and the oscilloscope's input. Further improvements using similar electronic devices are shown in Fig.~\ref{electronic} where the complete signal is faster than 25~ns. The final bias circuit scheme is shown in the lower panel of Fig.~\ref{electronic}.

To test which of the two SiPMs leads to a better time resolution, we used a Time to Digital Converter (TDC, CAEN V775N) with VME technology to measure the time that the light produced by a charged particle (from cosmic rays) 
took to pass from the scintillating plastic to the light sensor. Based on calibration, we converted the TDC units to time; one TDC unit (TDCU) is equivalent to 0.05~ns.

\begin{figure}[t]
\centering
\includegraphics[width=0.45\textwidth]{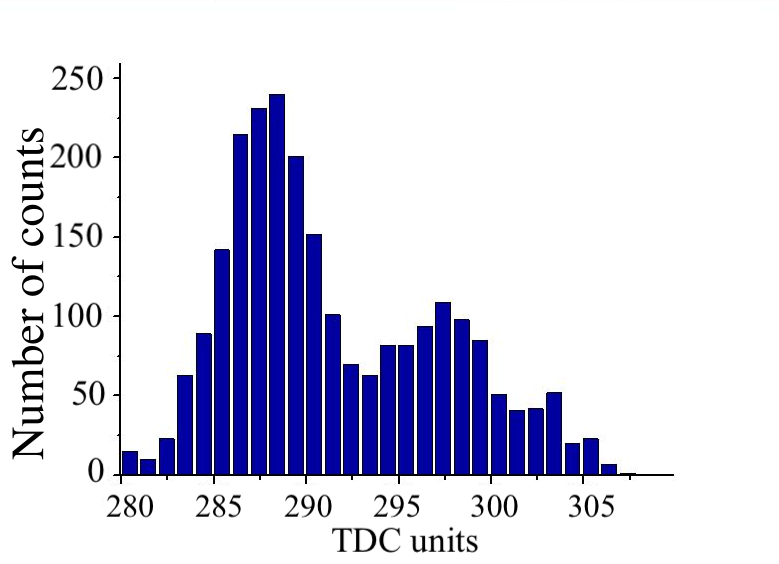}
\includegraphics[width=0.45\textwidth]{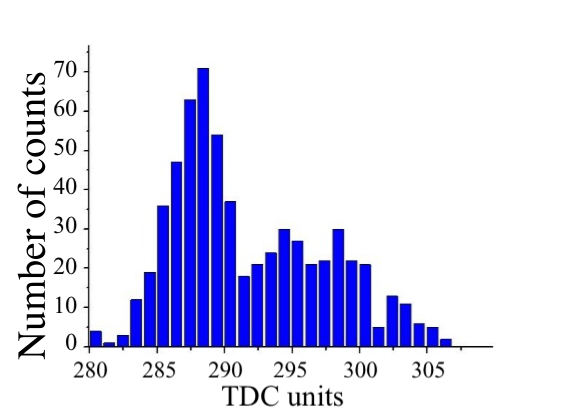}
\caption{Light acquisition distribution as a function of time in TDC units for C-60035-4P-EVB (top) and S10985 (bottom) SiPMs. The comparison shows a better time resolution for the S10985 SiPM which is around 2.4 TDCU = 120 ps.}\label{kindsensors}
\end{figure}

For the time resolution measurement, we used the BC404 scintillating material, which, according to our results, gave the best light acquisition. In addition, we used the two sensors (Hammamatsu and SensL) which were placed in three configurations on the scintillating plastic: both sensors in the middle of the same surface, one sensor on each surface, or both sensors at opposite sides. We observed that the
latter configuration gives the best light acquisition. Figure~\ref{kindsensors} shows the comparison between the output signals for the two sensors for this last configuration. The analysis was performed based on the light acquisition distribution obtained when using the telescope array. As shown in Fig.~\ref{kindsensors}, each of the signals consist of two peaks. This is due to the telescope-trigger used for the measurements. The procedure to clean up the signal is discussed further down in this section. From the analysis, we observe that the Hammamatsu SiPM signal is better than the SensL SiPM. The time resolution is obtained from the standard deviation of the time difference distribution between the trigger and the light sensor signal. 
A Gaussian fit to the first peak provides a better time resolution analysis. We infer that the SensL SiPM enables a signal with better statistics, albeit the Hamamatsu SiPM provides a better time resolution. The reason is that the peak for the BC404 emission spectrum (410~nm) is located near the Hamamatsu SiPM's maximum value for photo-detection efficiency at 450~nm \cite{hamma}.

\begin{figure}[t!]
\centering
\includegraphics[width=0.4\textwidth]{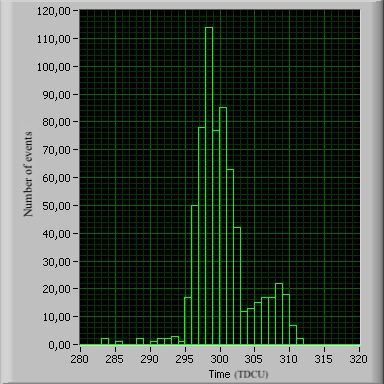}\\
\includegraphics[width=0.47\textwidth]{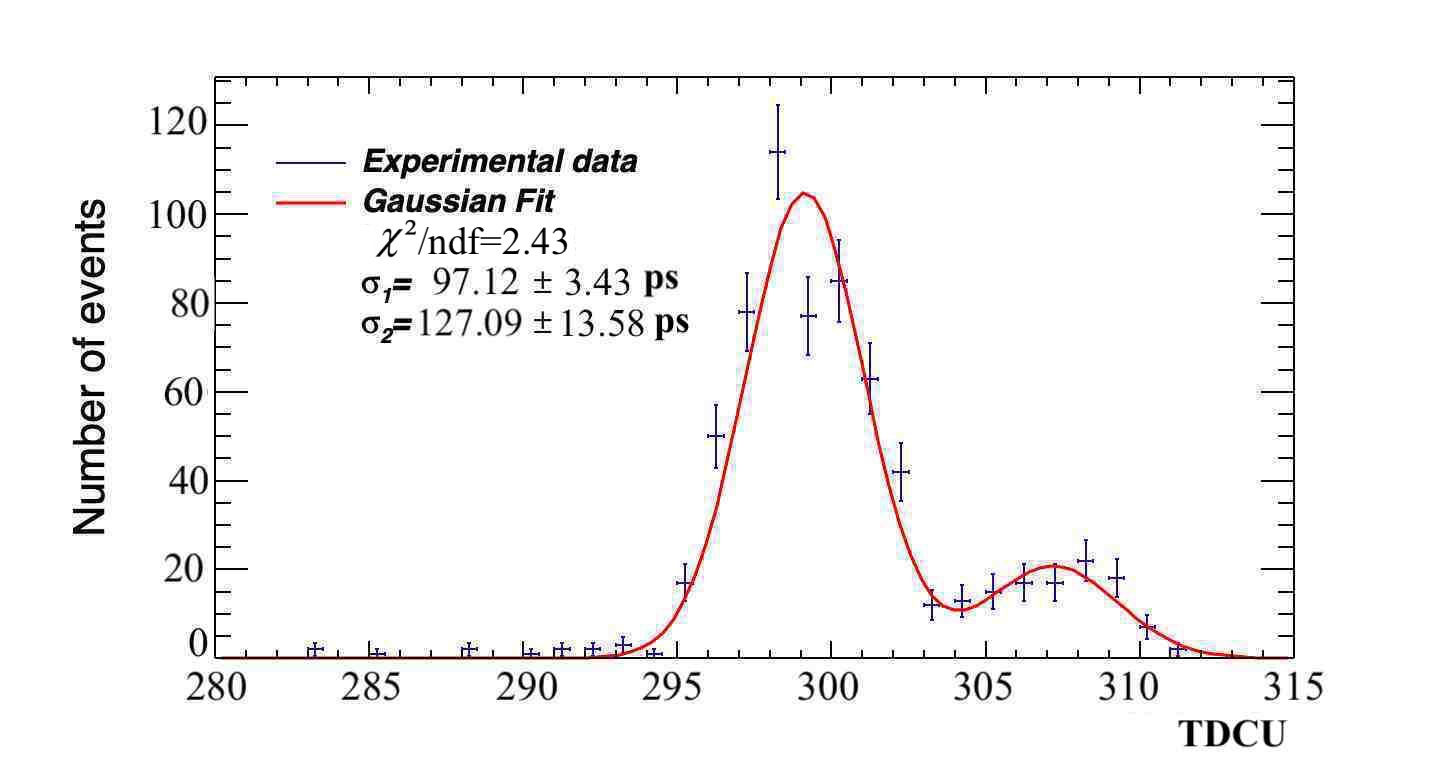}
\caption{Upper pannel: Light acquisition distribution as a function of the time difference  between the trigger and the SiPM's delay signal in TDCU. Lower pannel: Gaussian fit to the light acquisition distribution as a function of the time difference from where we obtained a time resolution around 97~ps. 
}\label{peaks}
\end{figure}

Next, we focused solely on the Hamamatsu S10985 SiPM and performed
further tests to improve the system’s time resolution. In this analysis, we used as a trigger the coincidence between one of the telescope’s
scintillating detectors and the BC404-S10985 system. When using only one reference scintillating plastic we were able to obtain better statistics. The measurement and the analysis fit for the system's time resolution are shown in Fig.~\ref{peaks}. We found a time resolution of 97.12$\pm$3.43~ps. 
\begin{figure}[b!]
\centering
\includegraphics[width=0.4\textwidth]{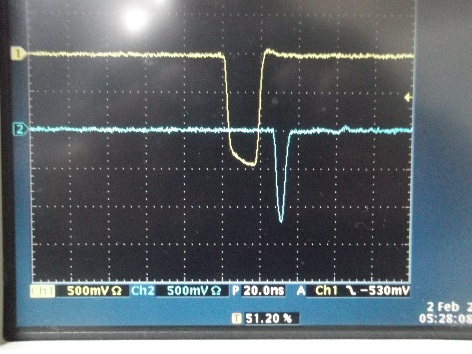}
\caption{The upper signal is the trigger when the two signals in coincidence were 4~ns wide. The lower signal shows the final width of the light sensor 
digital signal occurring during the trigger time.}\label{delayedsignal}
\end{figure}
In order to perform the analysis, we optimized the digital signals that define the trigger to make them finer. These signals were originally some 10~ns wide and produced a variation of the time difference that depended on the time interval when both trigger signals were in 
coincidence. This variation could be about 1 or 2 ns. The width of both  trigger signals was finally reduced to around 4~ns. 
These signals are illustrated in Fig.~\ref{delayedsignal}.

\begin{figure}[t]
\centering
\includegraphics[width=0.4\textwidth]{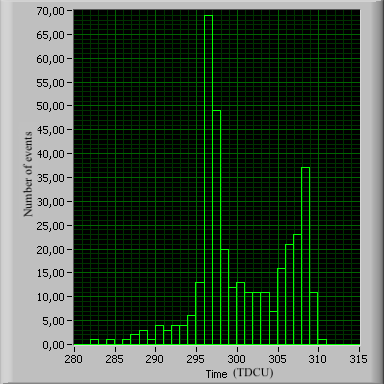}\\
\includegraphics[width=0.5\textwidth]{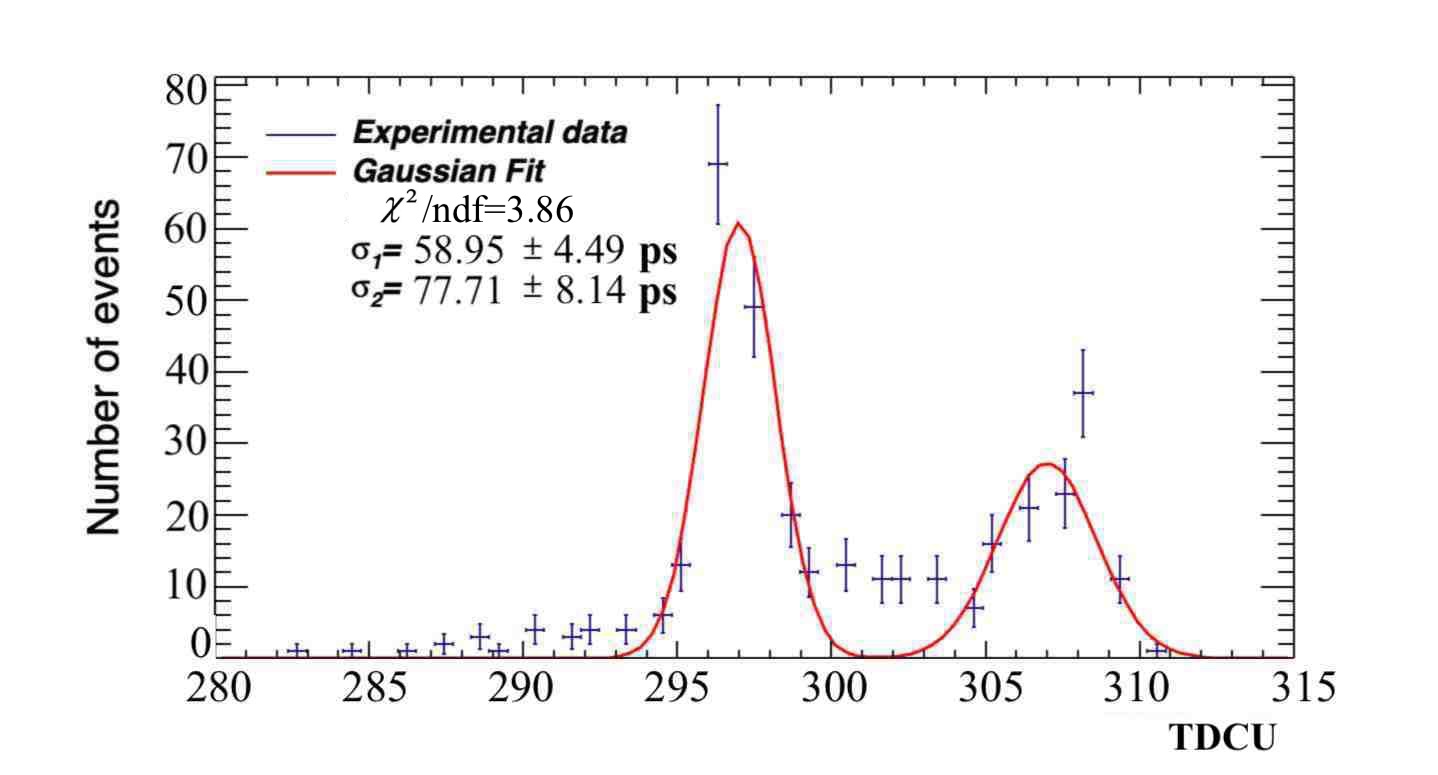}
\caption{Top: Light acquisition distribution as a function of time in TDCU, using the telescope array with the radioactive source near one sensor with the BC404 as the scintillating material. 
Bottom: Gaussian fits to the light acquisition distribution from where we obtained a standard deviation around 68~ps. Note the presence of the second peak (see discussion at the end of Sec.~\ref{3}). $\sigma_1$ and $\sigma_2$ are the sigma values of the fit for the first and second peak, respectively. }
\label{width}
\end{figure}

\begin{figure}[t]
\centering
\includegraphics[width=0.4\textwidth]{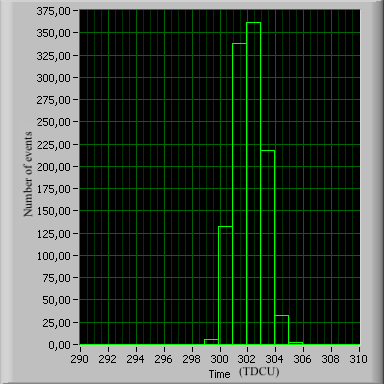}
\includegraphics[width=0.5\textwidth]{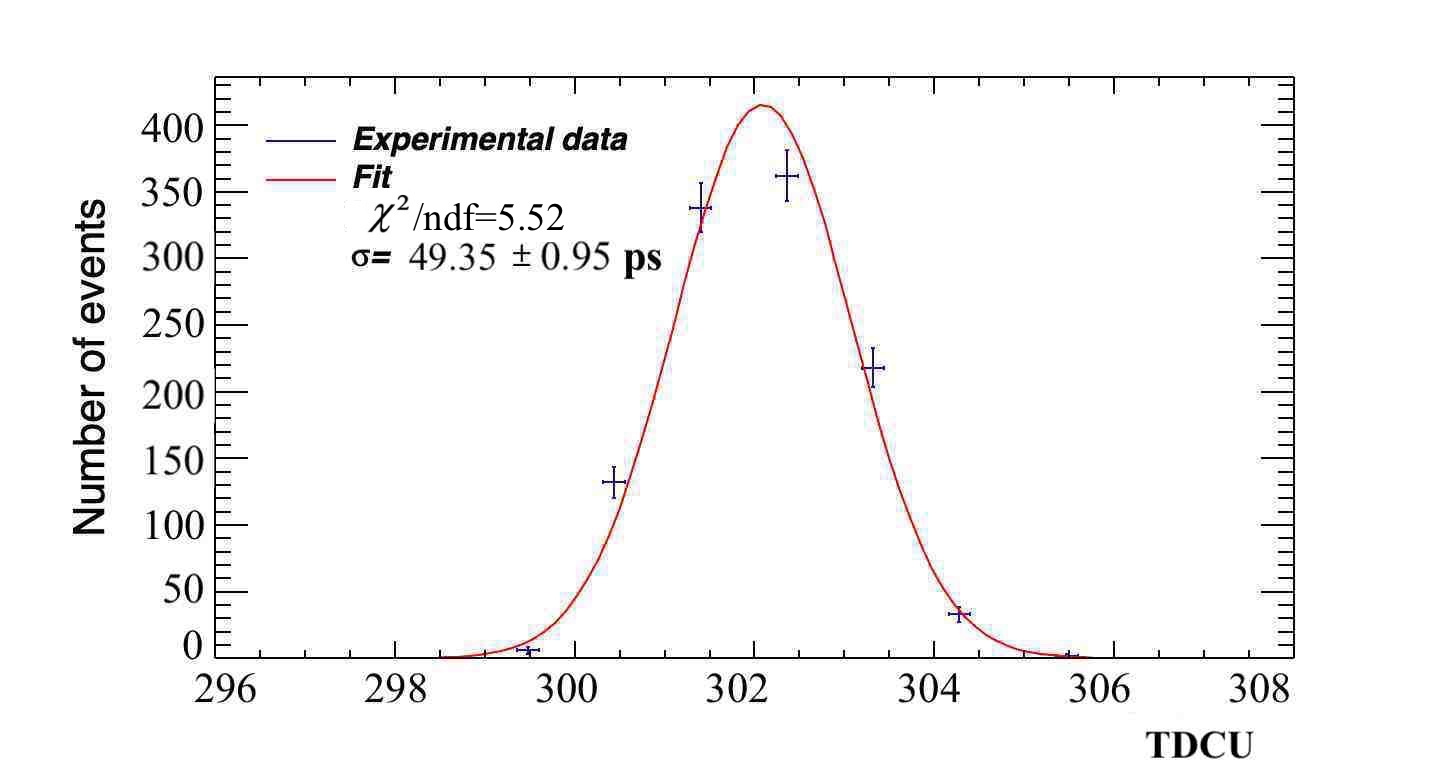}
\caption{Top: Light acquisition distribution as a function of the time difference between the light sensor trigger and the delay of the light sensor signal itself, in TDCU. Bottom: Gaussian fit to the light acquisition distribution from where we obtain a standard deviation $\sigma\simeq 49.35$~ps.}
\label{middle}
\end{figure}

For the last measurement, the source was placed near the surface where the sensors were located in both scintillating plastics.  Figure~\ref{width} shows the results of the analysis, which reveals a time resolution of 58.95$\pm$4.49~ps. Notice that the time resolution in this configuration is better than the one shown in Fig.~\ref{peaks}. This is expected since in this configuration, the produced photons arrive faster to the sensors than for the first configuration. To summarize our time resolution measurements:
\begin{itemize}
    \item For the configuration where the source is located in the middle of each scintillating plastic, the time resolution is 97.12$\pm$3.43~ps.
    \item For the configuration where the source is located near each sensor, the time resolution is 58.95$\pm$4.49~ps.
\end{itemize}
The second peak in the distribution of the number of counts as a function of time has been previously studied in Ref.~\cite{secondpeak}. In order to understand the nature of this second peak, we performed tests varying the trigger conditions. When triggering only with the sensor signal, that is, without using any coincidence, the histogram shown in Fig.~\ref{middle} was obtained, from where we observe that the second peak disappears and the time resolution amounts to 49.35$\pm$0.95~ps. Therefore, we attribute this second peak to the electronic trigger set up. This test gives also an estimate of the limits to the precision for the measurement of the time difference for the detection of the charged particles between the trigger and the tested scintillating plastic and sensor systems. 

\section{Discussion and conclusions}\label{4}

In this work we reported on tests made for three different scintillating plastics and two SiPMs to explore what configuration provides the best time resolution to detect fast moving charged subatomic particles. We found that out of the three tested plastics, the BC404 is the most efficient. The best time resolution was obtained with the Hammamatsu SiPM. For this set of scintillating plastic/SiPM combination, we performed the time resolution measurement for two configurations using a telescope trigger arrangement and a radioactive source of charged particles. We also observed that when using this trigger, it is sufficient to work with only one reference plastic. The two configurations for the time resolution measurement were: a) Sources located in the middle of each scintillating plastic and b) sources located next to each SiPM. We obtained time resolutions in the range from 58.95$\pm$4.49~ps to 97.12$\pm$3.43~ps. In addition, in order to figure out the origin of the second peak in the charge count distribution vs.\ time, we repeated the measurement without considering the telescope arrangement, which removes the second peak. In fact, with this configuration the time resolution improves to 49.35$\pm$0.95~ps.

Based on these results, we conclude that it is possible to obtain a time resolution below 50~ps, for example, using two SiPMs or reducing the dimensions of the scintillating plastics. These results can be used to improve the design for the SiPM distribution on the scintillating plastic surface to obtain the optimal signal acquisition for the MPD beam-beam counter detector. 
As shown in Ref.~\cite{NIM}, a time resolution around 45~ps can be achieved for the 5~cm heigh, 2~cm wide hexagonal tested plastic cell coupled to one SiPM. The results of this work agree with those of Ref.~\cite{NIM} when accounting both for the larger dimensions of the tested plastic cells and for the fact that lower energy charged particles, such as the ones obtained from the Na-22 radioactive source, are less efficiently detected~\cite{FFD}. Finally, we notice that the second peak in the charge distribution vs. time when using the telescope array is due to the set up electronic trigger. In order to get rid of this second peak, one can use a Constant Fraction Discriminator instead of the simple discriminator. The results of this work will be used for the final design and construction of a trigger system to be employed as part of MPD at NICA. 

\section*{Acknowledgments}
Support for this work has been received in part by Consejo Nacional de Ciencia y Tecnolog\'ia grant number 256494 and by UNAM-DGAPA-PAPIIT grant numbers IN107915 and IG100219. M.E.T-Y and L.V-C acknowledge the support provided by Universidad de Sonora in the initial stages of this work and the continuing support and commitment provided now by Universidad de Colima. M.R.C. thankfully acknowledges computer resources, technical advise and support provided by Laboratorio Nacional de Superc\'omputo del Sureste de
M\'exico (LNS), a member of the Consejo Nacional de Ciencia y Tecnolog\'ia network of national laboratories, with grant number No.53/Primavera 2017 and VIEP-BUAP grants 100524451-VIEP2018 and 100467555-VIEP2018.

\end{document}